%% file: stegamma.tex
\documentclass[aps,nofootinbib,preprint]{revtex4}

\usepackage{array}
\usepackage{graphicx}
\usepackage{amsmath}

\input mylatex

\newcommand{\met}{\not\!\!\!E_{T}}

\newcommand{\BK}[4]{\left<#1,#2|#3,#4\right>}
\newcommand{\BSK}[6]{\left<\!\,#1,#2|\not{\!#3}_{#4}|#5,#6\right>}
\newcommand{\BSSK}[8]{\left<#1,#2|\not{\!#3}_{#4}\not{\!#5}_{#6}|#7,#8\right>}
\newcommand{\fl}{\fcal_{L}}
\newcommand{\fr}{\fcal_{R}}
\newcommand{\g}{\gcal_{4f}}

\begin{document}

\begin{flushright}
hep-ph/0608331
\end{flushright}

\title{Search for new physics via single top production at TeV energy $e\gamma$ colliders}

\author{Qing-Hong Cao}
\email{qcao@ucr.edu}
\affiliation{Physics Department, University of California Riverside, Riverside, CA 92521, USA}

\author{Jose Wudka}
\email{jose.wudka@ucr.edu}
\affiliation{Physics Department, University of California Riverside, Riverside, CA 92521, USA}

\date{\today}

\begin{abstract}
We study the sensitivity of $ e \gamma$ colliders to physics beyond the \sm, when such
interactions are natural and their scale lies below the collider energy. Using the
reaction $ e \gamma \to b t \nu$ as a specific example, we show that the deviations
from the \sm\ can be parameterized in a model independent way by only 3 numbers. We
investigate the sensitivity of several observables to these 3 quantities, using the
various collider polarization modes to increase the signal to noise ratio.
These results are then translated into the expected sensitivity derived from this reaction 
to the new physics effects present in several specific models.
\end{abstract}

\maketitle

\section{introduction}

The top quark is by far the heaviest known fermion with a mass at the
electroweak symmetry breaking scale. Assuming this is not a
coincidence, it is hoped that a detailed
study of top quark couplings to other particles will be of utility
in clarifying whether the Standard Model (SM) provides the 
correct mechanism for electroweak
symmetry-breaking, or whether new physics is responsible.

It is therefore of interest to provide a general description 
of the top-quark couplings, which might be modified due to the
presence of new interactions and/or particles. In this
paper we will assume that such new effects are indeed
present, but that the energies available at present and near-future colliders lie
below the their typical new-physics  scale $\Lambda$.
In this case the characteristics of the new interactions can be probed only
through their virtual effects on processes involving SM 
particles; such effects can be efficiently coded in a 
model-independent way using the well-studied effective-Lagrangian
formalism~\cite{Georgi:1991ch,Weinberg:1978kz,Wudka:1994ny}.

The search for deviations from the SM couplings in single top
production has become one of the main focus in the forthcoming experiments
at the Large Hadron Collider (LHC) and the various Linear Collider
(LC) proposals. For example, the sensitivity to non-standard $Wtb$
couplings at the LHC via the single top quark production has been
studied in several papers~\cite{Kane:1991bg,Carlson:1994bg,Malkawi:1994tg,
Espriu:2001vj,Espriu:2002wx,Chen:2005vr,Batra:2006iq,Alwall:2006bx}. It has also
been shown~\cite{Weiglein:2004hn} that a very high energy LC with 
$500\,{\rm fb}^{-1}$ luminosity
will eventually improve the LHC limits by a factor of 3-8,
depending on the coupling under consideration.

Future linear colliders are expected to be designed to function
also as $ e\gamma$ and $\gamma\gamma$ colliders through
Compton back-scattering of laser light off one or both 
lepton beams~\cite{Ginzburg:1981vm}; in these modes the flexibility 
in polarizing both lepton and photon beams will
allow unique opportunities to analyze the top quark properties
and interactions. This is illustrated in \cite{Boos:1997rd,Cao:1998at}
for the case of the $Wtb$ coupling; and in \cite{Grzadkowski:1997cj}
for that of the four-Fermi operators in $e^+e^-\to t\bar t$ 
and top-quark decay.

It is generally the case that a given process receives contributions
from a variety of new-physics effects which are difficult to disentangle.
We will show that an exception to this is provided by single-top
production in an $e\gamma$ collider. We argue that in natural theories new physics
effects for this process can be parameterized by only three quantities,
and we will provide observables that can be used to measure them independently

This paper is organized as follows. In Sec.~\ref{sec:framework}, we list the
dominating new-physics contributions to the single-top
production in $e\gamma$ collisions and summarize the existing experimental
constraints on the corresponding coefficients. In Sect.~\ref{sec:phenomenological}
we analyze the effects of the non-SM coefficients in the distribution of
several observables as function of the beam energy and polarization parameters.
These results are then used to study the sensibility of this reaction to
the effects predicted by several models (Sect.~\ref{sec:distinguishing}). 
Conclusions and parting comments are presented in Sect.~\ref{sec:conclusions}.

\section{Framework}
\label{sec:framework}

In this section, we summarize the basic elements of the framework
used in the analysis (additional details are included in the appendices).

\subsection{Effective Lagrangian}

The  arguments suggesting the existence of physics beyond the SM
and the uses of the effective Lagrangian
approach (ELA) for describing these new interactions have been extensively 
discussed~\cite{Georgi:1991ch,Weinberg:1978kz,Wudka:1994ny}.
In the ELA it is assumed that none of the heavy excitations can be
directly produced, so that all new physics effects can be parameterized
by gauge-invariant operators of dimension higher than four constructed out
of the SM fields. These higher-order operators are suppressed by inverse powers
of the new physics scale $\Lambda$ (the scale at which the excitations
of the underlying theory can be directly probed). Among the effective
operators those of dimension 5 necessarily violate lepton 
number~\cite{Weinberg:1979sa,Wilczek:1979hc,Weldon:1980gi},
and are strongly bounded by existing data~\cite{Buchmuller:1985jz};
the largest contributions are then expected to be generated by dimension-6
operators, which we denote as $\ocal_{i}$ 
The effective Lagrangian then takes the form~\footnote{It is worth mentioning 
that the effects of some tree-level
induced dimension-8 operators may compete with those from  dimension-6
operators if the latter are generated at the one-loop level, and thus have associated
a suppression factor of $1/(4\pi)^{2}$.} 
\beq
\lcal_{eff}=\lcal_{SM}+\frac{1}{\Lambda^2}
\sum_i\left(C_i\ocal_i+ \hbox{H.c.} \right)+O\left(\Lambda^{-3}\right),
\eeq
where $C_i$ are coefficients that parametrize the non-standard
interactions. Some of the $\ocal_i$ can be generated by tree
graphs in the underlying theory, while other are 
necessarily loop-generated~\cite{Arzt:1994gp}
and the corresponding $C_i$ will be suppressed by a numerical factor $\sim1/16\pi^{2}$. We
will therefore focus our attention on the tree-level induced operators
only, and examine their effects in the single top production in $e\gamma$
collisions. These operators fall in two categories: those modifying 
the $Wtb$ coupling, and those that generate four fermion interactions; we will
discuss them separately

There are only 2 tree-level generated operators of the first type:
\bea
\ocal_{\phi q}\up3 & = & i\left(\phi^{\dagger}\tau^{I}D_{\mu}\phi\right)\left(\bar{q}\gamma^{\mu}\tau^{I}q\right)+ \hbox{H.c.} ,\\
\ocal_{\phi\phi} & = & i\left(\phi^{\dagger}\epsilon D_{\mu}\phi\right)\left(\bar{t}\gamma^{\mu}b\right)+ \hbox{H.c.} ,
\eea
where $\phi$ denotes the SM scalar doublet, $D_{\mu}$ the covariant
derivative, $q(\ell)$ the quark (lepton) isodoublets and
$t(b)$ the corresponding isosinglets
(we follow the notation of~\cite{Buchmuller:1985jz}).
After symmetry breaking, these operators generate the
following contribution to the $Wtb$ coupling
\beq
\lcal_{Wtb}\up{\rm dim-6}=\frac{g}{\sqrt{2}}\left\{ \bar{t}\gamma^{\mu}\left(\fl P_{L}+\ \fr P_{R}\right)bW_{\mu}^{+}+ \hbox{H.c.} \right\} ,
\label{eq:wtb}
\eeq
with 
\beq
\fl = \frac{C_{\phi q}^{\left(3\right)}v^{2}}{\Lambda^{2}},\qquad
\fr = \frac{C_{\phi\phi}v^{2}}{2\Lambda^{2}},
\label{eq:flfrdef}
\eeq
where $v=246\,{\rm {\rm GeV}}$ is the vacuum expectation value (VEV). 

There are 4 operators of the second type: 
\beq
\begin{array}{lcl}
\ocal_{qde}  = \left(\bar{\ell}e\right)\left(\bar bq\right), & &
\ocal_{\ell q} =  \left(\bar{\ell}e\right)\epsilon\left(\bar{q}t\right),\cr
\ocal_{\ell q^{\prime}}  = \left(\bar{\ell}t\right)\epsilon\left(\bar{q}e\right), & &
\ocal_{\ell q}^{(3)} = \frac{1}{2}\left(\bar{\ell}\gamma_{\mu}\tau^{I}\ell\right)\left(\bar{q}\gamma^{\mu}\tau^{I}q\right).
\end{array}
\eeq
all of which can be generated at tree level~\cite{Arzt:1994gp}.
The first three, however, involve a chirality flip;  in a  natural
theory this implies that the corresponding coefficients will be
proportional to $m_{e} $ and can be ignored. Hence,
we only need to consider the last operator, $\ocal_{\ell q}^{(3)}$,
from which we extract out the following effective $\ell\nu bt$ vertex:
\beq
\lcal_{4f} = \frac{\g}{\Lambda^{2}}\left\{ \left(\bar{\nu}\gamma^{\mu}P_{L}e\right)
\left(\bar b\gamma_{\mu}P_{L}t\right)+\left(\bar{e}\gamma^{\mu}P_{L}\nu\right)\left(\bar{t}\gamma_{\mu}P_{L}b\right)\right\} ,
\label{eq:4fop}
\eeq
with $\g =  C_{\ell q}\up3/2$.

In our calculation we will take all the effective couplings to be
real in order to simplify our analysis. We will also
assume that the $\nu\ell W$ vertex does not receive significant 
contributions from physics beyond the SM~\footnote{A rough estimate
shows that the scale of new physics that would modify this
vertex lies above $\sim 7\,\tev$.}.

\subsection{Constraints of the effective operators}

The LEP precision data requires $\left|\fl\right|\leq0.02$~\cite{Larios:1999au},
assuming no deviations from the SM $ttZ$ vertex. Recent data on $b\to s\gamma$ 
provides the limit $\left|\fr\right|<0.004$~\cite{Chetyrkin:1996vx,Larios:1999au,Burdman:1999fw},
provided one neglects other possible new-physics effects, such
as those embodied in a $bstt$ 4-fermion interaction (that could be generated, 
for example, by a heavy $W'$ vector boson with flavor-changin couplings).

The four-fermion interaction operator will violate the unitarity constraints
at high energies; the resulting constraint (see, for 
example~\cite{Gounaris:1996yp}) is
\beq
\left|\g\right|<\frac{16\pi}{s}\Lambda^{2} < 16\pi.
\eeq
since the ELA is valid only for $ \Lambda > s $.
We note that even when $\g = O(50)$ the 4-fermion operators in Eq.~\ref{eq:4fop} will
only affect the branching ratio for $t\to b\ell\nu_{\ell}$ at the $0.1\%$ 
level~\footnote{This is is because the SM contributions to the amplitude peak in the
region of phase space where $\left(p_{\ell}+p_{\nu}\right)^{2}\simeq M_{W}^{2}$
in which case their interference with the $\g$ term can be ignored; the
new-physics corrections to the differential decay rate are then $O(1/\Lambda^{4})$.}, 
well within the current experimental bound $Br(t\to b\ell\nu)=9.4\pm2.4\%$~\cite{Eidelman:2004wy}.
In contrast the effects of this operator in single top quark production
can be very significant. This is because this reaction is dominated
by the $t$-channel processes involving a virtual $W^*$ boson ($q_{W}^{2}<0$)
which receives two kinematic enhancements: \emph{(i)} the $W^*$ 
propagator does not suffer the $1/s$ suppression%
\footnote{The invariant mass of the virtual $W$ boson peaks around $200\,{\rm GeV}$
for $\sqrt{s}=500\,{\rm GeV}$, and at $300\,{\rm GeV}$ for $\sqrt{s}=1\,{\rm TeV}$%
}, and \emph{(ii)}  the photon-splitting ($\gamma\to b\bar b$) collinear
enhancement. Therefore the interference effects of the SM contribution
and the $\g$ contribution can be large.

Due to the strong constraint on $\fr$ (assuming no cancellations with
a possible $bstt$ contribution), its effects are negligible. Hence,
we will concentrate on the effects of couplings $\fl$ and $\g$ 
in various kinematics distributions (though, for completeness, we will include some
effects generated by $ \fr$). It is clear, however, that $\fl$ merely produces a change
in the overall normalization of the SM cross-section, so it will be
difficult to disentangle its contribution from the SM background unless
one can measure the event rate very accurately. 

\section{Phenomenological study}
\label{sec:phenomenological}

The expressions for the cross section and helicity amplitudes for the process
of $e\gamma\to \nu t b$ are given in the appendices. Using these
results we examine the sensitivity of this reaction to the effective
operators mentioned above, and discuss how beam polarization can 
be used to optimize this sensitivity. 

For the numerical evaluation we choose the following set of SM input
parameters: $\alpha=1/137.0359895$, $G_{\mu}=1.16637\times10^{-5}\,{\rm GeV^{-2}}$,
$M_{W}=80.35\,{\rm GeV}$, $\Gamma_{W}=2.0887\,{\rm GeV}$, $M_{Z}=91.1867\,{\rm GeV}$,
$m_{e}=0.51099907\,{\rm MeV}$. The square of the weak gauge coupling
is then $g^{2}=4\sqrt{2}M_{W}^{2}G_{\mu}$ and the branching ratio
of the $W$ boson into leptons is $Br(W\to\ell^+\nu)=0.108$~\cite{Cao:2004yy}
(including the $O(\alpha_{s})$ corrections to $W\to\bar{q}q^{\prime}$).
 
At the linear collider it will be possible to adjust the initial-state
electron and positron longitudinal-polarizations $\pcal_{e}$
and $\pcal_{\tilde{e}}$, the average helicities of the initial-state
photons $\pcal_{\gamma}$ and $\pcal_{\tilde{\gamma}}$,
and their maximum average linear-polarization $\pcal_{t}$ and
$\pcal_{\tilde{t}}$ with the azimuthal angles $\varphi$ and
$\tilde{\varphi}$ (we use the same conventions as~\cite{Ginzburg:1981vm}).
In this study we restrict ourselves to the 8 choices 
shown in Table~\ref{tab:polarizationparameters}.
The first (last) 4 sets correspond to circularly (linearly)-polarized
initial photons; the spin-density matrix depends on $\varphi$ only for the
last two sets (see Appendix~\ref{sec:Photon-distribution-functions}); for these 
cases we found that the cross section is maximized when $\varphi\sim 1.18$, 
so we will use this value for the rest of calculation.

\begin{table}

\caption{The choices of the polarization parameters.\label{tab:polarizationparameters}}

\begin{tabular}{>{\centering}p{0.6in}|>{\centering}p{0.5in}>{\centering}p{0.5in}>{\centering}p{0.5in}|>{\centering}m{0.6in}|>{\centering}p{0.5in}>{\centering}p{0.5in}>{\centering}p{0.5in}}
\hline 
 & $\pcal_{e}$ & $\pcal_{t}$ & $\pcal_{\gamma}$&& $\pcal_{e}$ & $\pcal_{t}$ & $\pcal_{\gamma}$ \tabularnewline \hline 
(1) & $+1$ & $0$ & $+1$ & (5) & $+1$ & $+1$ & $0$ \tabularnewline \hline 
(2) & $-1$ & $0$ & $+1$ & (6) & $-1$ & $+1$ & $0$ \tabularnewline \hline 
(3) & $+1$ & $0$ & $-1$ & (7) & $+1$ & $1/\sqrt{2}$ & $1\sqrt{2}$ \tabularnewline \hline 
(4) & $-1$ & $0$ & $-1$ & (8) & $-1$ & $1/\sqrt{2}$ & $1/\sqrt{2}$ \tabularnewline \hline
\end{tabular}
\end{table}

\subsection{Inclusive cross section}

\begin{figure}
\includegraphics[clip,scale=0.6]{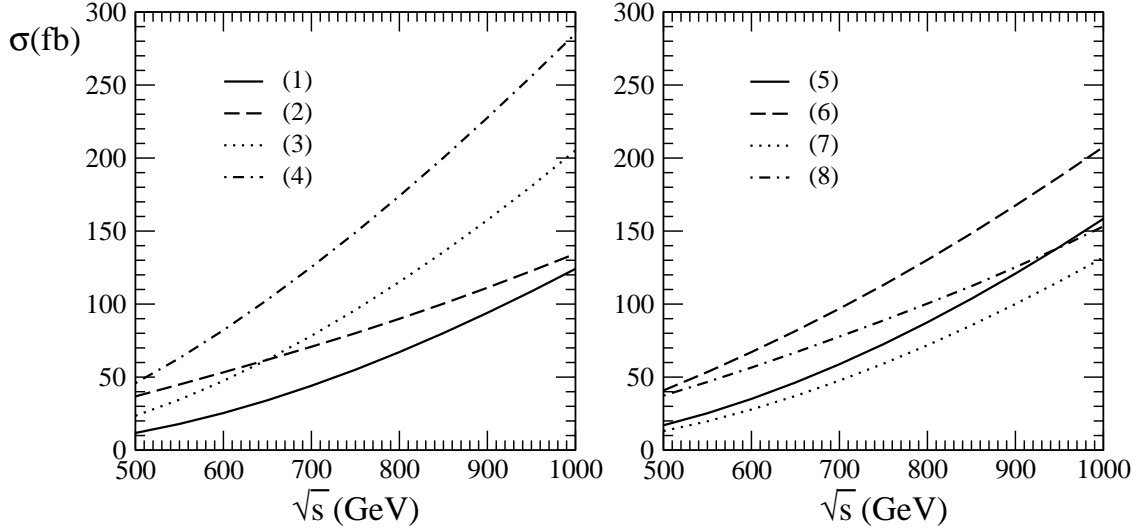}
\caption{The inclusive cross section of the process $e^{+}\gamma\to\bar{\nu}t\bar b$
as a function of the beam energy in the SM. The labels (1-8) denote
the choice of polarization parameter set defined in the text.
\label{fig:sm-xsec}}
\end{figure}

In Fig.~\ref{fig:sm-xsec} we show the inclusive cross section of
the process $e^{+}\gamma\to\bar{\nu}t\bar b$ as a function of the
beam energy ($\sqrt{s}$) for the various choices of polarization
parameters (PP). We note that the cross section is significantly larger
for set (4), which is due to the fact that in this case almost all
the scattered photons have negative helicity (see Fig.~\ref{fig:photonPDF}),
so their interactions with the right-handed positrons are unsuppressed
by factors of the electron mass~\footnote{In contrast, for set (1) 
almost all scattered photons will have positive helicity leading to an 
$m_e$ suppression factor.}~\cite{Muhlleitner:2005pr}.

Including the contributions of the effective operators, the cross
section takes the following form 
\bea
\sigma & = & \sigma_{SM}+\sum_{i}c_{i}\delta\sigma_{i}+\cdots  \cr
 & = & \sigma_{SM}+\fl\delta\sigma_{\fl}+\fr\delta\sigma_{\fr}+
\g' \delta\sigma_{\g'}+\cdots
\label{eq:xsec}
\eea
where 
\beq
\g' = \g \left(\frac{1{\rm TeV}}\Lambda \right)^2,
\label{eq:gp}
\eeq
$\sigma_{SM}$ is the SM cross section, and the $\delta\sigma_{i}$
are the non-standard contributions generated by the effective operators.
In this section we will consider one parameter at a time with the
other couplings being fixed to the SM values; since we assume $s\ll\Lambda^2$
we can ignore all the terms of order of $1/\Lambda^{4}$. 

The terms containing $\fl$ are proportional to the SM contributions
so $\delta\sigma_{\fl} = 2 \sigma_{SM}$. The remaining $\delta\sigma_i$
are presented in Table~\ref{tab:xsec-ratio} 
for various choices of polarizations and beam energies. We find that 
$ \delta \sigma_{\g}\sim 0.1 \sigma_{SM} $ so that a few thousand events
should be sufficient to probe $ \g'$ to $O(1)$

{\eightrm
\begin{table}

\caption{The various contributions to the cross section for the reaction
$e^{+}\gamma\to\bar{\nu}t\bar b$ (\ref{eq:xsec}) with the various
choices of the polarization parameters in table \ref{tab:polarizationparameters}.}
\label{tab:xsec-ratio}
\begin{tabular}{|c|c|c|c|c|c|c|c|c||c|c|c|c|c|c|c|c|}
\hline 
$\sqrt{s}$ (\tev) &
\multicolumn{8}{c||}{0.5} & \multicolumn{8}{c|}{1.0}  \cr\hline\hline 
{\rm PP set} &
(1) & (2) & (3) & (4) & (5) & (6) & (7) & (8) &
(1) & (2) & (3) & (4) & (5) & (6) & (7) & (8) \cr\hline
$\sigma_{SM}$(fb) &
23.4 & 73.6 & 47.0 & 91.6 & 34.0 & 81.6 & 26.2 & 74.6 &
248 & 280 & 412 & 570 & 324 & 432 & 270 & 318  \cr\hline
$\delta\sigma_{\fr}/\sigma_{SM}$ &
0.45 & 0.18 & 0.25 & 0.15 & 0.35 & 0.15 & 0.43 & 0.18 &
0.03 & 0.23 & 0.15 & 0.14 & 0.20 & 0.15 & 0.20 & 0.25 \cr\hline
$\delta\sigma_{\g'}/\sigma_{SM}$ &
0.09 & 0.12 & 0.10 & 0.12 & 0.10 & 0.12 & 0.09 & 0.12 &
0.11 & 0.11 & 0.12 & 0.12 & 0.12 & 0.12 & 0.11 & 0.11 \cr\hline
\end{tabular}
\end{table}
}

\subsection{Basic kinematic distributions}

The relatively large effects observed in the total cross section suggest that 
other observables might be useful in probing the details of the effective
interactions by choosing the PP set that maximizes
their significance $\scal/\sqrt{\bcal}$ of the observables under consideration
($\scal$ denotes the signal and $\bcal$ the SM background). 
Below we follow this procedure for several kinematic variables
(in most of the cases we examined the PP set (2) is preferred). 

Specifically, given a variable $ \phi $ we expand the differential 
cross section $ d\sigma /  d\phi $ in terms of the effective operator
coefficients as in (\ref{eq:xsec}) and define the normalized
distribution functions
\beq
f_{SM}(\phi) = \inv{\sigma_{SM}} \frac{d\sigma_{SM}}{d\phi}, \qquad
f_{4f}(\phi) = \inv{\delta\sigma_{\g'}} \frac{d\delta\sigma_{\g'}}{d\phi}.
\label{eq:disf}
\eeq
The values of these quantities, combined with the results of 
table \ref{tab:xsec-ratio}, can be used to determine the usefulness
of a given choice of $ \phi$ in observing or bounding the magnitude
of the new physics effects.

The top quark produced via single-top quark process is highly polarized
due to the nature of left- or right-handed charged weak current 
interaction~\cite{Cao:2004ky,Cao:2004ap,Cao:2005pq}.
Hence a strong spin correlation exists between the final state particles
and the initial state leptons. In order to fully understand these
correlations, we first examine the ideal case where the back-scattered
photon is either left-handed or right-handed polarized. We then 
study the spin correlation effects with a realistic photon beam. In
the last part of this section we examine the effective operators effects
in the distribution of several other kinematic variables.

In the following we will choose $ +\hat z$ as the direction of the 
incoming photon; we define $\theta_t,~\theta_{\bar b}$ as the polar angles 
of the $t$ and $\bar b$ quarks respectively and denote $\theta_{t\bar b}$ as
the angle between the directions of the $t$ and $\bar b$ quarks

\subsubsection{Angular distributions for perfectly polarized photon beams}

\begin{figure}
\includegraphics[scale=0.6]{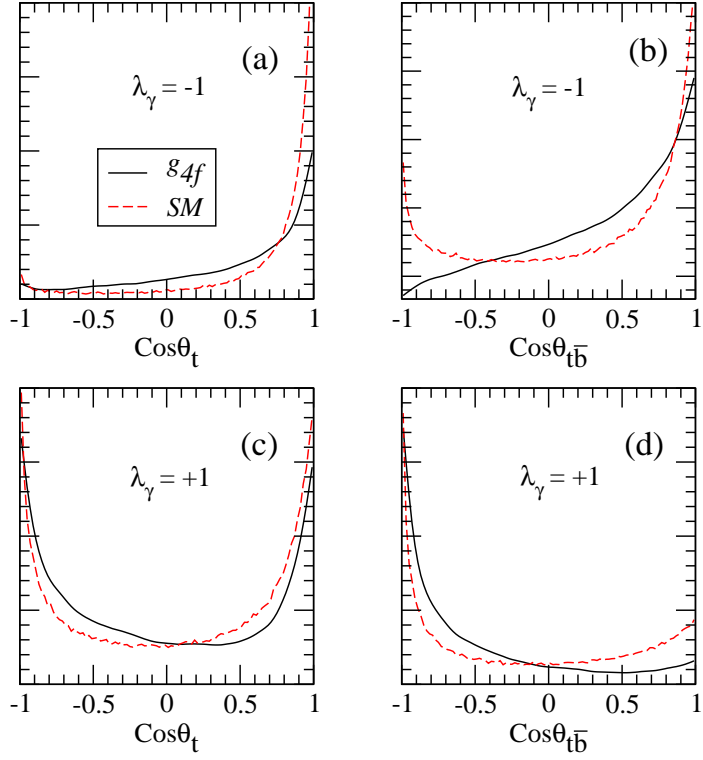}
\caption{Normalized distributions (\ref{eq:disf}) for
$\phi=\cos\theta_t$ and $\phi=\cos\theta_{t\bar b}$
for purely left ($\lambda_{\gamma}=-1$) and right-handed photons
($\lambda_{\gamma}=+1$), and $\sqrt{s}=500\,{\rm GeV}$. Solid curves: 
$f_{4f}$; dashed curves: $f_{SM}$.}
\label{fig:cth_100polarized}
\end{figure}

For photon energies $\gg m_b$ the final-state $\bar b$ quarks
move preferentially along the beam line direction 
due to the collinear enhancement, hence 
$\cos\theta_{\bar b}\simeq1$ irrespective of the photon polarization.
In contrast both $\cos\theta_t$ and $\cos \theta_{t\bar b} $ are
very sensitive to the polarization of the incoming photon; in order
to fully understand the differences between these quantities
we examine (Fig.~\ref{fig:cth_100polarized}) their distributions for 
purely polarized beams in the center of momentum frame.

Within the SM, left-handed photons ($\lambda_{\gamma}=-1$) preferentially
generate top and $\bar b$ quarks moving parallel or anti-parallel
to the incoming photons ($\cos\theta_t,~\pm\cos\theta_{t\bar b} \simeq1$).
In contrast, right-handed photons generate top quarks moving either
parallel or anti-parallel to the direction of the incoming photon,
with the $\bar b$ moving in the opposite direction ($\cos\theta_t,
~-\cos\theta_{t\bar b}\simeq\pm1$).
These distributions remain essentially unchanged when the four-Fermion
contribution is included except when $\lambda_\gamma = -1$,
in which case the $\bar b$ will preferentially move parallel to the $t$. 

\begin{figure}
\includegraphics[scale=0.7]{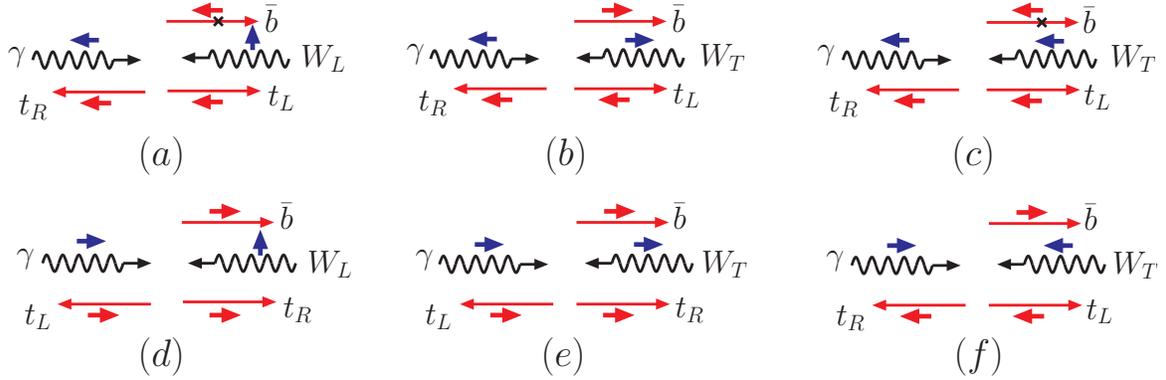}
\caption{Kinematic configurations of the $t$ and $\bar b$ quarks when both
of them move along the beam line. The upper (lower) three plots correspond
to left (right)-handed polarized photons. The long straight (waved)
lines denote the the direction of motion of the fermions (bosons).
The short bold arrows denote the particle's spin direction; a cross
on a fermion line indicates a mass insertion, which flips the fermion's
helicity.}
\label{fig:spin-kinematics}
\end{figure}

The SM results can explained within the effective-$W$ boson 
approximation (EWA)~\cite{Dawson:1984gx}, that has been used in the study of heavy
quark and lepton production~\cite{Dawson:1986tc,Yuan:1987mk}. The
method is based on the observation that at high energies the $W^{\pm}$
and $Z^{0}$ bosons can be treated as parton constituents of the leptons,
in which case a $t$-channel single top quark event in hadron collisions
is dominated by diagrams containing a longitudinal $W^*$
boson~\cite{Yuan:1989tc} since its couplings to the fermions are enhanced
by powers of $m_{t}/m_{W}$ relative to those of the transversely polarized
bosons~\footnote{Though the separation into transverse and
longitudinal components is not Lorentz invariant, the transverse degrees
of freedom remain transverse under boosts in the beam-line direction;
though the longitudinal ones mix with the temporal ones, gauge invariance
of the physical amplitude for the sub-process insures that the correct
result is preserved.}. For the same reason the SM contributions to the 
process $e^+\gamma\to\bar\nu t\bar b$ are dominated by $t$-channel-like 
exchange diagrams Figs.~\ref{fig:feyndiag_wtb_em}(a),(b). The main features of
Fig.~\ref{fig:cth_100polarized} then follow from angular momentum
conservation, see Fig.~\ref{fig:spin-kinematics}.

These considerations cannot be extended to the case where the 4-fermion
operator (Eq.~\ref{eq:4fop}) is included since such an operator
can be interpreted as being produced by the exchange of a heavy charged $W'$ gauge
boson in the underlying theory, whose typical energy lies {\em below}
its mass, so that the EWA approximation cannot be used, and
the contribution from the lepton ($e^+\bar\nu$) and heavy-quark 
($t\bar b$) lines cannot be factorized. Nevertheless, since the 
main deviations from the SM are produced by the the interference
effects between the SM and the 4-fermion operator, many of the
features of the SM are observed in the $f_{4f}$ distributions
for $\cos \theta_t$ and $\cos\theta_{t\bar{b}}$.

\subsubsection{Angular distributions for realistic photon beams.}

For photon beams generated through Compton back-scattering,
the distributions for $\theta_t,~\theta_{\bar b}$ and $\theta_{t\bar b}$
exhibit roughly the same behavior as for the previous ideal case.
In particular $\cos\theta_t$ remains sensitive to both the effective
operators and initial-state polarization parameters. 

\begin{figure}
\includegraphics[scale=0.6]{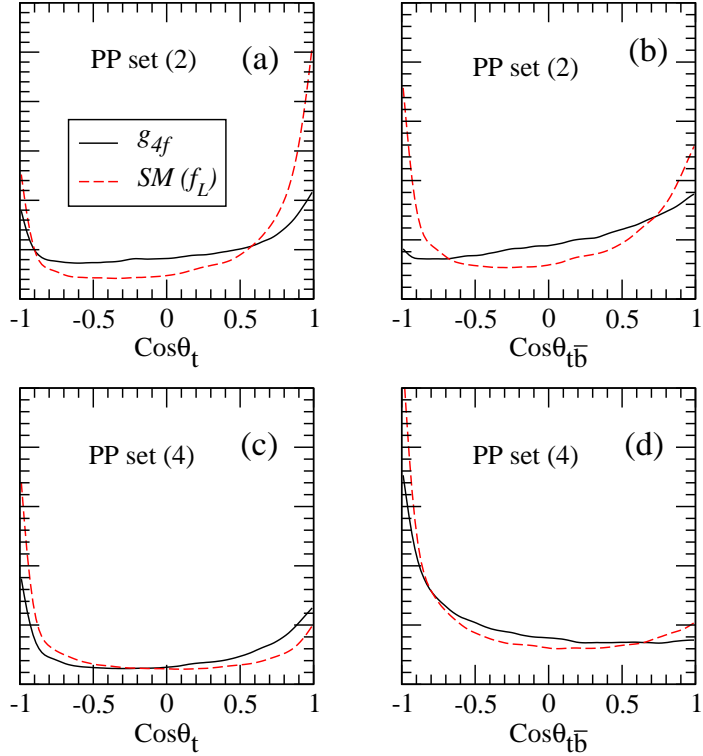}
\caption{
Same as in fig. \ref{fig:cth_100polarized} for realistic photon beams.
Figures (a) and (b) were obtained using the polarization parameter
set (2); (c) and (d) using set (4). \label{fig:ctht}}
\end{figure}

In Figs.~\ref{fig:ctht} we show the distributions of $\cos\theta_{t}$
and $\cos\theta_{t\bar b}$ for the PP sets (2) and (4) for $\sqrt{s}=500\,{\rm {\rm GeV}}$.
From this it is clear that left-handed polarized photons dominate
for PP set (2), which is consistent with the photon spectrum distributions
in Fig.~\ref{fig:photonPDF}(2). As a result, the SM contribution
peaks when the top quark moves along the beam line in the forward
direction, while the $\g$ contribution shifts the top quark off the
beam axis, (Fig.~\ref{fig:ctht}(a)). The net contribution of left
and right-handed photons results in two peaks in the SM contribution
to the $\cos\theta_{t\bar b}$ distribution, while the corresponding
$\g$ contribution is flat (Fig.~\ref{fig:ctht}(b)).

The origin of the distributions for PP set (4) is less direct. Though
the spectrum of the left-handed polarized photon dominates over the
one of right-handed photon (Fig.~\ref{fig:photonPDF}(4)), the sub-process
amplitudes of the right-handed photon are enhanced by the contribution
of the longitudinal polarized $W$ boson. The distributions in Fig.~\ref{fig:ctht}(c)
and (d) result from a combination of these effects.

\subsubsection{Missing energy}

\begin{figure}
\includegraphics[scale=0.6]{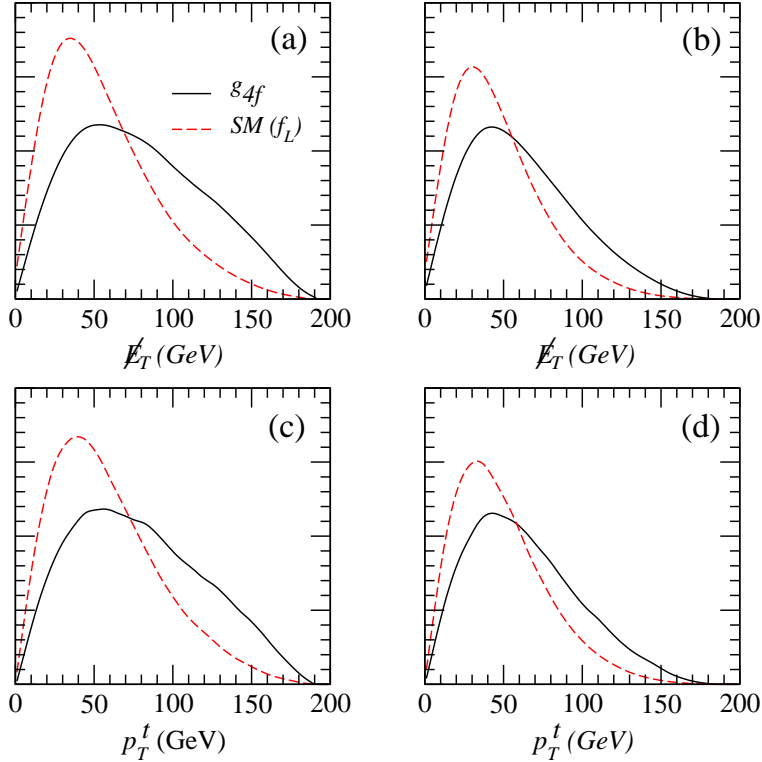}
\caption{Normalized distributions (\ref{eq:disf}) $f_{SM}$ (dashed curves)
and $f_{4f}$ (solid curves) for 
the missing energy ($\met$) and the top-quark
transverse momentum ($p_{T}^{t}$), for $\sqrt{s}=500\,{\rm GeV}$.
(a) and (c) PP set 2, (b) and (d) PP set 4.
\label{fig:pt}}
\end{figure}

Figs.~\ref{fig:pt}(a) and (b) show the normalized distributions
(Eq.~\ref{eq:disf}) for
the missing energy ($\met$) carried by the final-state neutrino.
Within the SM the neutrino comes from the initial state positron after
emitting a $W$ boson, so its transverse momentum peaks at $\sim M_{W}/2$.
The $\g$ contribution tends to shift the missing energy to the large
transverse momentum region, which can be understood if we assume
the four-Fermi operator is induced by a heavy $W'$ boson: due to
the large $W'$ mass, the neutrino produced by a virtual $W'$ boson will
have a larger $p_{T}$. 

\subsubsection{Rapidity of the bottom quark}

\begin{figure}
\includegraphics[scale=0.5]{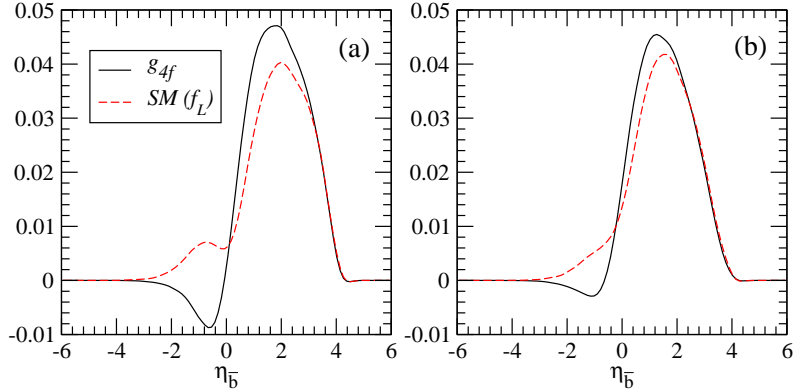}
\caption{The normalized distribution of the rapidity of the anti-bottom quark
for $\sqrt{s}=500\,{\rm GeV}$: (a) PP set (2), (b) PP set (4).  The red dashed curves denote the SM
contributions; solid black curves correspond to the interference of
the $\g$ and SM graphs .\label{fig:yb}}
\end{figure}

Fig.~\ref{fig:yb} shows the rapidity distribution of the bottom
quark. Since the bottom quark is predominately produced from the initial
state photon splitting, (Fig.~\ref{fig:feyndiag_wtb_em}(b)), its
rapidity peaks in the forward direction, (dashed curves in Fig.~\ref{fig:yb}).
The $WW\gamma$ diagram (Fig.~\ref{fig:feyndiag_wtb_em}(d)) corresponds
to a virtual $W$ boson moving in the negative rapidity region, balancing
the $\bar{\nu}$ emitted from the incoming $e^{+}$. This virtual
$W$ boson has a large invariant mass and it is produced mainly in
the central rapidity region; its decay products, the $\bar b$ and
$t$ quarks, also populate this region, which leads to the small kink
in the $\eta_{\bar b}\sim -1$ region. The $\g$ contribution
slightly shifts the bottom quark distributions towards the central region
where it generates a negative peak. This reduces the SM kink at
$\eta_{\bar b} \sim -1 $ and enhances the peak in the forward region.

\section{Phenomenological consequences.}
\label{sec:distinguishing}

In this section we discuss the accuracy with which effective couplings
$\fl$ and $\g$ can be measured in the reaction $e^{+}\gamma\to\bar{\nu}t\bar b$
process for beam energies of $\sqrt{s}=500\,{\rm GeV}$ and $1\,{\rm TeV}$.
Using these results we will discuss the extent to which various models
can be differentiated.

The behavior observed in figure \ref{fig:pt} suggests that
that a cut on the missing energy
or transverse top-quark momentum, $p_T^t,~\met\geq 60 \gev$ will suppress 
the SM contribution and
allow an accurate measure the effective coupling $\g$. Using the
results of tables \ref{tab:xsec-ratio} and \ref{tab:indf} together with
(Eq.~\ref{eq:gp}) and (Eq.~\ref{eq:disf}) we find that with a luminosity
of $100\,\rm{fb}^{-1}$ the four-femrion interaction will produce more 
than a 3-standard deviation effect for $ \g' > 0.3$; the
corresponding numbers for $ 500\,\rm{fb}^{-1} $ and $ 1000\,\rm{fb}^{-1}$ are
$ \g' > 0.13$ and $ \g' > 0.09$ respectively. These results might be modified
by possible correlations with $ \fl,~\fr$; in order to include
these effects we performed an optimal-observable analysis, to which we
now turn.
\begin{table}
\caption{Integrated distribution functions for the neutrino missing energy and 
top-quark transverse momentum with $ \sqrt{s} = 500 \gev$}
\label{tab:indf}
$$
\begin{array}{|c|l|l|l|}
\hline
PP & \phi  & \int_{60}^\infty d\phi f_{4f}(\phi) & \int_{60}^\infty d\phi f_{SM}(\phi) \cr
\hline\hline
2  & \met  & \quad 0.62                          & \quad 0.38                          \cr
   & p_T^t & \quad 0.63                          & \quad 0.42                          \cr
\hline
4  & \met  & \quad 0.47                          & \quad 0.28                          \cr
   & p_T^t & \quad 0.49                          & \quad 0.31                          \cr
\hline
\end{array}
$$
\end{table}

\subsection{Optimal observable analysis}

The optimal observable technique is a useful tool for estimating expected
statistical uncertainties in various coupling measurements. Suppose
we obtain a differential cross section in terms of some convenient
kinematic variables $\phi$, 
\beq
\Sigma(\phi) = \frac{d\sigma}{d\phi} =\sum_{i}c_{i}f_{i}\left(\phi\right)\label{eq:cs}
\eeq
where $f_{i}\left(\phi\right)$ are known functions and $c_{i}$ are
model-dependent coefficients; the goal is to determine the accuracy
to which the $c_{i}$ can be measured.

Denoting the statistical uncertainty in $c_{i}$ by $\Delta c_{i}$,
the optimized covariance matrix equals 
\beq
V_{ij} = \vevof{\Delta c_i \,\Delta c_ j } =\frac{\sigma_{T}}N X_{ij},
\eeq
where $\sigma_{T} = \int\Sigma(\phi) d\phi$, $N$ is
the total number of events and $X_{ij}$ is obtained from
\beq
\left(X^{-1}\right)_{ij} = \int\frac{f_{i}(\phi)f_{j}(\phi)}{\Sigma(\phi)}d\phi.
\label{eq:matrix-M}
\eeq
In particular, the minimum uncertainty in the measurement of $c_{i}$
is given by
\beq
\overline{\Delta c_{i}}=\sqrt{\vevof{\left(\Delta c_{i})^{2}\right)} }
=\sqrt{\frac{\sigma_{T}}N X_{ii}}
\label{eq:deltac}
\eeq

To order of $1/\Lambda^{2}$ we can write
\beq
\frac{d\sigma}{d\phi}=f_{SM}(\phi) + \fl f_{\fl}(\phi) + \fr f_{\fr}(\phi) + \g' f_{\g}(\phi).
\label{eq:weighting}
\eeq
where $\g'$ is defined in (\ref{eq:gp}).
To this order we need only calculate $\left(X^{-1}\right)_{ij}$ with
$i,j=1,2,3,4$ corresponding to SM, $\fl$, $\fr$ and $\g$, respectively,
and then obtain $\overline{\Delta c_{i}}$ using (\ref{eq:deltac})
taking $\sigma_{T}=\int d\phi\, f_{SM}\left(\phi\right)$. 

In Table~\ref{tab:optimalobs} we present the choice of PP and kinematic
variable that minimize the statistical uncertainty for the effective
operator parameters. We note that the best choice for $\fr$ corresponds
to PP sets (1) or (3) that enlarge the right-handed polarized photon
beam.

\begin{table}

\caption{The minimal statistical errors $\overline{\Delta c_{i}}$ ($c_{i}=\fl$, $\fr$
, $\g$) for various energies and luminosities (we only include the PP set giving the lowest error)}
\label{tab:optimalobs}

\def\oldtable{
\begin{tabular}{|c|c|c|c|c|c|c|c|c|c|c|}
\hline 
& $\phi$ & PP &
\multicolumn{3}{c|}{$\sqrt{s}=500\,{\rm TeV}$} &
\multicolumn{1}{c|}{$\phi$} &
PP & \multicolumn{3}{c|}{$\sqrt{s}=1\,{\rm TeV}$} \cr \hline 
 & & &
$100\,{\rm fb}^{-1}$ & $500\,{\rm fb}^{-1}$ & $1000\,{\rm fb}^{-1}$ &
 & & $100\,{\rm fb}^{-1}$ & $500\,{\rm fb}^{-1}$ &
$1000\,{\rm fb}^{-1}$\cr \hline 
$\overline{\Delta\fl}$ & $\cos\theta_{te}$ & 8 & 0.10 & 0.046 & 0.033 & $\cos\theta_{te}$ &
4 & 0.036 & 0.016 & 0.011 \cr \hline 
$\overline{\Delta\fr}$ & $\cos\theta_{te}$ & 3 & 0.12 & 0.054 & 0.038 & $p_{T}^{b}$ &
1 & 0.026 & 0.012 & 0.0083 \cr\hline 
$\overline{\Delta\g'}$ & $\met$ & 6 & 0.13 & 0.059 & 0.042 & $p_{T}^{b}$ & 4 & 0.049 &
0.022 & 0.016 \cr\hline
\end{tabular}
}

$$
\begin{array}{|c|c|c|c|c|c|}
\multicolumn{6}{c}{\sqrt{s}=0.5\,{\rm TeV}} \cr\hline
\overline{\Delta c_i} & \phi  	    & PP & 100\,{\rm fb}^{-1} & 500\,{\rm fb}^{-1} & 1000\,{\rm fb}^{-1} \cr \hline\hline
\overline{\Delta\fl}  & \cos\theta_{te} & 8  & 0.10 			& 0.046 		   & 0.033 			 \cr \hline 
\overline{\Delta\fr}  & \cos\theta_{te} & 3  & 0.12 			& 0.054 		   & 0.038 			 \cr \hline 
\overline{\Delta\g'}  & \met 		    & 6  & 0.13 			& 0.059 		   & 0.042 			 \cr \hline
\end{array}
\quad
\begin{array}{|c|c|c|c|c|c|}
\multicolumn{6}{c}{\sqrt{s}=1\,{\rm TeV}} \cr\hline
\overline{\Delta c_i} & \phi 		    & PP & 100\,{\rm fb}^{-1} & 500\,{\rm fb}^{-1} & 1000\,{\rm fb}^{-1} \cr\hline\hline
\overline{\Delta\fl}  & \cos\theta_{te} & 4  & 0.036 			& 0.016 		   & 0.011 			 \cr\hline 
\overline{\Delta\fr}  & p_{T}^{b}       & 1  & 0.026 			& 0.012 		   & 0.0083 		 \cr\hline 
\overline{\Delta\g'}  & p_{T}^{b} 	    & 4  & 0.049 			& 0.022 		   & 0.016 			 \cr\hline
\end{array}
$$
\end{table}

These result indicate the possibility of measuring these couplings at the 5\%\ accuracy 
(at the $1\sigma$ level) in a $ \sqrt{s} = 0.5~\tev $ collider with a luminosity of $500~\hbox{fb}^{-1}$.
If we denote by $ \Lambda_{\rm max}$ the largest scale that can be probed with a
given measurement then, using the expressions for $ \fl,~\fr$ (Eq.~\ref{eq:flfrdef}) with $C_i\sim1$, 
this corresponds to a reach  up to $\Lambda_{\rm max} = 1~\tev$; using $\g$ this 
becomes $\Lambda_{\rm max} = 5~\tev$.
For $ \sqrt{s} = 1~\tev $ and a luminosity of $1000~\hbox{fb}^{-1}$
the corresponding values are $\Lambda_{\rm max} = 1~\tev$ and $10~\tev$ respectively.

The types of physics being probed are quite different: a bound on $ \fl, \fr$
constrains a combination of the mixing angles of the $t$ and $b$ with heavy generations and  
those of the $W$ with a heavy $W'$ and the masses of these heavy excitations; 
a bound on $\g$ constrains a combination of fermion couplings to heavy $W'$ and/or leptoquarks
and the masses of these particles.

\subsection{Distinguishing models}

As mentioned above the effective operators contributing to the reaction
$e\gamma\to t b \nu$ can be generated by a variety of heavy particles.
Below we consider several models:

\begin{itemize}
\item Normal $W'$ model:\\
An extra gauge boson $W^{\prime}$ arises in many extension
of the SM~\cite{Pati:1974yy,Mohapatra:1974hk,Mohapatra:1974gc,
Senjanovic:1975rk,Georgi:1989xz,Georgi:1989ic,Li:1981nk,Chivukula:1995gu,
Malkawi:1996fs,Muller:1996dj,He:1999vp,Datta:2000gm}.
Here we only consider the simplest case where the
new gauge boson ($W'$) has the same couplings as the SM
$W$ boson. If we also assume no $W-W'$ mixing 
the low-energy $W'$ effects correspond to
\beq
\fl =  0,\quad
\g' = -\frac{g^{2}}{2} \left( \frac{1~{\rm TeV}}\Lambda \right)^2 ,\quad
\Lambda = m_{W'}.
\eeq
When $m_{W'}=1\,{\rm TeV}$, $\g=0.21$.
\item Hexagonal $SU(3)$ unification model~\cite{Cao:2005ba}\\
In this supersymmetric model the original $SU(3)^{6}$ symmetry
is broken to $\su2_A\times\su2_B$ at a high scale which break to
$\su2$ at the SUSY braking scale. As a result a second set of weak gauge
bosons appear at this stage,
with the coupling of the $W'$ boson to the SM particles is completely
fixed by the underlying of theory; the salient feature of this model
is that the $W'$ boson couples to the leptons and quarks differently.
In general, the SM $W$ boson can mix with $W'$ boson but for illustration,
we consider the simplest case where there is no $W-W'$ mixing; then
\beq
\fl = 0,\quad
\g' = -\frac{g^2}{2}\frac{g_A}{g_B}
\left(\frac{g_A}{g_B}\cos^2\theta-\frac{g_B}{g_A}\sin^2\theta\right)
\left( \frac{1~{\rm TeV}}\Lambda \right)^2 ,\quad \Lambda = m_{W'}.
\eeq
where $g_A= 0.71,~g_B=1.63$ are the $\su2_{A,B}$ gauge couplings 
(fixed by the unification condition), and $\theta$ is
the mixing angle that determines the coupling of the $W'$ to the
leptons (it depends only on $M_{W'}/M_W$). When $m_{W'}^{2}=1\,{\rm TeV}$, 
$\theta\simeq0.4$ and $\g'=0.05$.

\item Little Higgs Model with T-parity~\cite{Low:2004xc,Hubisz:2004ft,Hubisz:2005tx}\\
The LH model can be extended to include a discrete symmetry, T-parity,
which greatly reduces the new-physics contributions to precision electroweak
observables~\cite{Hubisz:2005tx}. In particular, light and heavy
gauge bosons have opposite charges under T-parity and do not mix;
the four-fermion operator in Eq.~\ref{eq:4fop} 
is not generated (at tree level) for the
same reason. Nonetheless the $Wtb$ is modified by the mixing of the
top quark with its T-even partner is still present and shifts the
coupling sizablely through the mixing with the SM top quark. 
In our notation, 
\beq
\fl = -\frac{c_{\lambda}^{4}}{2}\frac{v^{2}}{f^{2}},\quad
\fr = 0,\quad
\g' = 0, \quad
\Lambda = 4\pi f
\eeq
where $c_{\lambda}=\lambda_{1}/\sqrt{\lambda_{1}^{2}+\lambda_{2}^{2}}$.
Here $\lambda_{1,2}$ denote the Yukawa couplings giving masses to the
top quark and its heavy partners, and $ 4 \pi f$ the heavy symmetry breaking
scale. In this study, we choose $c_{\lambda}=1/\sqrt{2}$, $\fl=0.015~(0.007) $ for
$f=0.7~(1)\,{\rm TeV}$.
\end{itemize}

\begin{figure}
\includegraphics[scale=0.5]{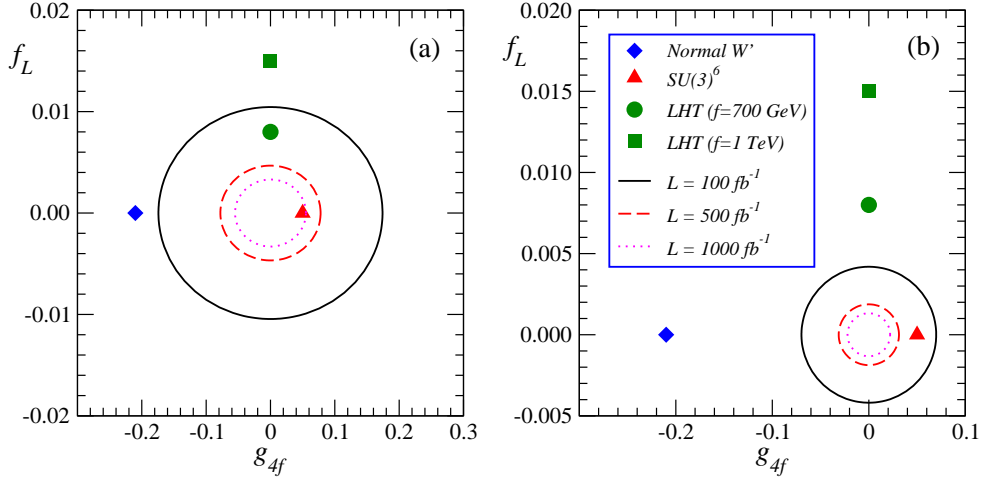}
\caption{$2\sigma$ contours of the total cross section
within the models studied in the text (a) $\sqrt{s}=500\,{\rm GeV}$,
(b) $\sqrt{s}=1\,{\rm TeV}$ (assuming $ m_{W'} = f = 1$ TeV and $\fr=0 $).}
\label{fig:2sigma-bouds}
\end{figure}

The $2\sigma$ bounds of the anomalous couplings parameter space,
within which no distinction from the SM is possible, are shown in
Fig.~\ref{fig:2sigma-bouds}. In the figure we also show the theory
predictions of the three models described above. It is clear that
the single top production process in the $e\gamma$ collision can
be used to distinguish various models which include either $\fl$
or $\g$ effect coupling (the $\fr$ coupling is not included
since the error of its measurement is much larger than its value).

\section{Conclusions}
\label{sec:conclusions}

In this paper we have considered single-top production in $e\gamma$
colliders as a probe for new physics effects, assuming that the 
collider energy is not sufficiently high to directly probe these
effects. We argued that for natural theories the deviations from
the SM tree-level couplings in this reaction can be parameterized by
3 couplings; one of these ($\fr$) is very strongly constrained
by the data while another ($\fl$) affects only the overall cross-section
normalization.

Of the observables studied the missing energy, transverse
top-quark momentum and scattering angle proved best for distinguishing the
effects of the various effective couplings. An optimal
observable analysis (table \ref{tab:optimalobs}) shows 
that these coefficients can
be measured with 3-$\sigma$ accuracy
down to the 10-5\%\ level depending on the
collider energy and luminosity; for a luminosity of
$ 1000~\hbox{fb}^{-1} $ this allows to probe physics at
scales up to 1.5 times the collider center mass energy. Even the
use of a broad observable such as the total cross section
can be used to probe new physics effects at similar levels.

These results, however, will be diluted when realistic
detector effects are included in the analysis. 
It is also noteworthy that the deviations from the SM do not show
up as unexpected peaks but as an excess or deficiency of evens in
various kinematic regions which can be reliably determined only
if the details of the detectors are well understood and sufficiently
large number of events has been accrued. Still we
believe that the above features are of sufficient interest
to warrant further study of this reaction.

\begin{acknowledgments}
Q.-H. Cao thanks C.-P. Yuan for many useful discussions.
Q.-H. Cao and J. Wudka are supported in part by the U.S. Department
of Energy under grant No. DE-FG03-94ER40837. 
\end{acknowledgments}

\appendix

\section{Photon distribution functions
\label{sec:Photon-distribution-functions}}

The photon distribution functions have been given with details in
\cite{Grzadkowski:2003tf}. For the sake of completeness, we
also list them here. 

At a linear collider the single top quarks can be produced from the
following two processes: 
\beq
t : ~ e^{+}\gamma\to\bar{\nu}t\bar b,\qquad
\bar{t}  : ~ e^{-}\gamma\to\nu b\bar{t},\label{eq:process}
\eeq
where the photon comes from the original incoming electron and positron,
respectively. The total cross sections for these two processes at
a linear collider are
\bea
\sigma\left(e^{-}e^{+}\rightarrow\bar{\nu}t\bar b\right) & = & \int d x\,\frac{ d N(x)}{ d x}\hat{\sigma}\left(e^{+}\gamma\rightarrow\bar{\nu}t\bar b\right),\cr
\sigma\left(e^{-}e^{+}\rightarrow\nu b\bar{t}\right) & = & \int d y\,\frac{ d N(y)}{ d y}\hat{\sigma}\left(e^{-}\gamma\rightarrow\nu b\bar{t}\right),
\eea
 where $ d N/ d y$ is the photon-spectral function 
(equivalent to the  parton distribution function inside hadrons). Here
$x$ ($y$) denotes the fraction of the electron (positron)
momentum carried by the photons: $p_{\gamma}=xp_{e^{-}}$
and $p_{\gamma}=yp_{e^{+}}$. 

Using the spin density matrix method to keep track on the photon polarization,
the hard cross section the processes (\ref{eq:process}) take the form
\bea
\hat{\sigma}\left(e^+\gamma\to\bar\nu t\bar b\right) & = & \half N_c\sum_{a,b=\pm}\inv{2\hat s}\int d \Pi_3\rho^{ab}(x)
\mcal_a\left(e^+\gamma\to\bar\nu t\bar b\right)\mcal_{b}^{\star}\left(e^+\gamma\to\bar\nu t\bar b \right),\label{eq:partonlevel_t}\\
\hat{\sigma}\left(e^-\gamma\to\nu b\bar t\right) & = & \half N_c\sum_{a,b=\pm}\inv{2\hat s}\int d \Pi_3\rho^{ab}(y)
\mcal_{a}\left(e^-\gamma\to\nu b\bar t\right)\mcal_{b}^{\star}\left(e^-\gamma\to\nu b\bar t\right),\label{eq:partonlevel_tbar}
\eea
where the factor of $1/2$ comes from 
averaging over the initial-state electron or positron spin (it
should be replaced by $1$ for the purely polarized electron or positron beam),
$N_{c}=3$ is the color factor by summing over the color of the final
state quarks, $ d \Pi_{3}$ represents 3-body final-state phase
space, and $\rho^{ab}$ is the photon polarization density matrices;
the helicity amplitudes $\mcal$ are given in the next section. 
Here, $\hat{s}$ is the square of the
sub-process beam energy: $\hat{s}=\left(p_{e^{+}}+p_{\gamma}\right)^{2}$
for $e^{+}\gamma$ collision or $\hat{s}=\left(p_{e^{-}}+p_{\gamma}\right)^{2}$
for $e^{-}\gamma$ collision. The spin-averaging factor for the incoming
photon has been included in the spin density matrices $\rho^{ab}$.

The density matrix $\rho$ is not only a function of the momentum
fraction $x$ or $y$, but also depends on the choice of the photon
polarization vectors. 
In $e^+-\gamma$ collisions we take the positron moving along the
$-\hat{z}$ direction and the photon scattered from the incoming
electron along the $+\hat{z}$ direction, therefore, its
polarization vectors are defined as:
\beq
\epsilon^{\mu}\left(h_{1}\right)=\frac{1}{\sqrt{2}}\left(\epsilon_{1}^{\mu}+ih_{1}\epsilon_{2}^{\mu}\right),
\eeq
where $h_{1}$ is the helicity of the photon with $\epsilon_{1}^{\mu}=\left(0,1,0,0\right)$
and $\epsilon_{2}^{\mu}=\left(0,0,1,0\right)$. 

In $e^-\gamma$ collisions we take the
electron moving along the $+\hat{z}$ direction; the photon is
scattered off the incoming positron and moves along $-\hat{z}$
direction, therefore its polarization vectors are defined as
\beq
\epsilon^{\mu}\left(h_{2}\right)=\frac{1}{\sqrt{2}}\left(-\epsilon_{1}^{\mu}+ih_{2}\epsilon_{2}^{\mu}\right),
\eeq
where $h_{2}$ is the helicity of the photon. 

\begin{figure}
\includegraphics[scale=0.7]{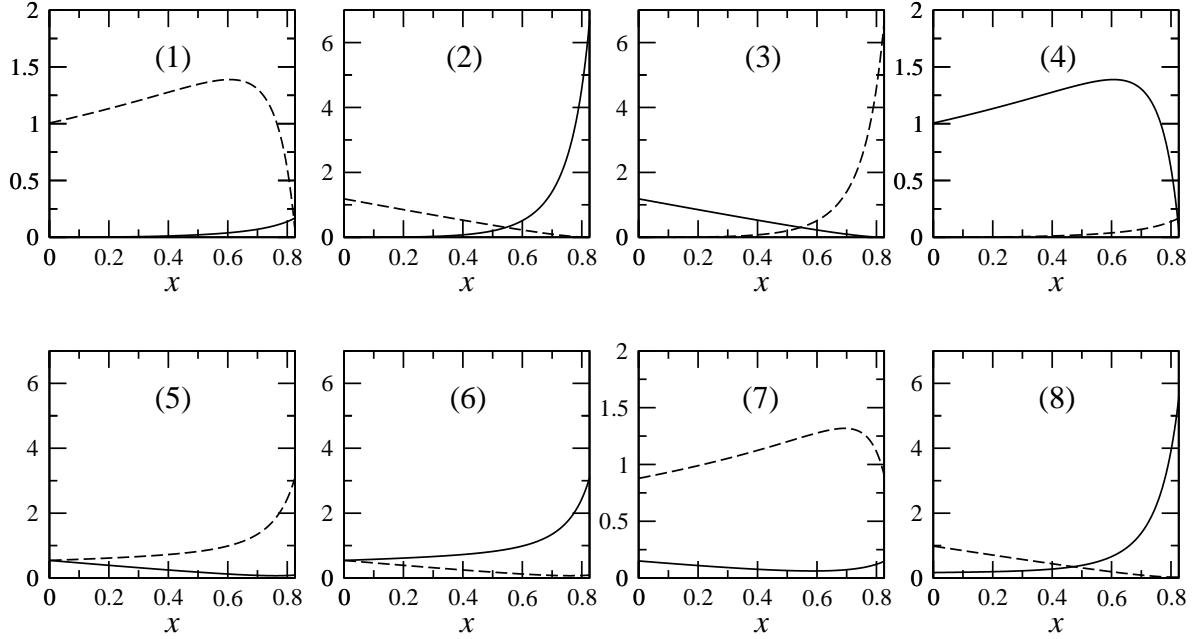}

\caption{The photon spectrum distributions as a function of $x$ for various
choices of the polarization parameters listed in Table.~\ref{tab:polarizationparameters}.
The solid curve in each plot denotes the spectrum of negative helicity
photons ($\rho\left(2,2\right)$)while the dashed curve denotes that
of positive helicity photons $\rho\left(1,1\right)$. \label{fig:photonPDF}}
\end{figure}

With these choices the density matrices $\rho(x)$
and $\rho(y)$ in terms of the three Stokes parameters $\xi_i$ are given by
\bea
\rho\left(x\right) & = & \frac{1}{2}\left(\begin{array}{cc}
1+\xi_{2}\left(x\right) & \xi_{3}\left(x\right)-i\xi_{1}\left(x\right)\\
\xi_{3}\left(x\right)+i\xi_{1}\left(x\right) & 1-\xi_{2}\left(x\right)\end{array}\right),\label{eq:spin-density-matrix-em}\\
\rho\left(y\right) & = & \frac{1}{2}\left(\begin{array}{cc}
1-\xi_{2}\left(y\right) & \xi_{3}\left(y\right)+i\xi_{1}\left(y\right)\\
\xi_{3}\left(y\right)-i\xi_{1}\left(y\right) & 1+\xi_{2}\left(y\right)\end{array}\right).\label{eq:spin-density-matrix-ep}
\eea
The maximum of $x$ and $y$ is given by 
\beq
x_{max} = \frac{x_{0}}{1+x_{0}},\qquad
y_{max} = \frac{y_{0}}{1+y_{0}}.
\eeq
Here $x_{0}$ ($y_{0}$) determines the upper limit of the final photon
energy,
\beq
E_{\gamma}^{max}=\frac{x_{0}}{1+x_{0}}E_{e}.
\eeq
The larger the value for $x_{0}$ is chosen the more energy can be
transferred to the photon. However, to suppress 
pair production we must require
\beq
x_{0}(y_{0})\leq2\left(1+\sqrt{2}\right)\sim4.828.
\eeq
The photon-spectrum function $ d N(x)/ d x$ and $\xi_{i}\left(x\right)$
in the spin density matrix $\rho\left(x\right)$ immediately after
its production at the conversion point are given by the following
formulas:
\bea
 &  & \frac{ d N\left(x\right)}{ d x}=\frac{C\left(x\right)}{D\left(x_{0}\right)},\label{eq:photon-spectrum}\cr
 &  & \xi_{1}\left(x\right)=\frac{2\pcal_{t}\sin\left(2\varphi\right)\left[r\left(x\right)^{2}\right]}{C\left(y\right)},\cr
 &  & \xi_{2}\left(x\right)=\frac{\pcal_{e}\, f_{2}\left(x\right)+\pcal_{\gamma}\, f_{3}\left(x\right)}{C\left(y\right)},\cr
 &  & \xi_{3}\left(x\right)=\frac{2\pcal_{t}\cos\left(2\varphi\right)\left[r\left(x\right)^{2}\right]}{C\left(x\right)},
\eea
where
\bea
 &  & C\left(x\right)=f_{0}\left(y\right)+\pcal_{e}\ \pcal_{\gamma}f_{1}\left(x\right),\cr
 &  & D\left(x_{0}\right)=D_{0}\left(x_{0}\right)+\pcal_{e}\pcal_{\gamma}D_{1}\left(x_{0}\right),\cr
 &  & f_{0}\left(x\right)=\frac{1}{1-y}+1-x-4r\left(x\right)\left[1-r\left(x\right)\right],\cr
 &  & f_{1}\left(x\right)=\frac{y\left(2-x\right)\left[1-2r\left(x\right)\right]}{1-y},\cr
 &  & f_{2}\left(x\right)=y_{0}r\left(x\right)\left\{ 1+\left(1-x\right)\left[1-2r\left(x\right)\right]^{2}\right\} ,\cr
 &  & f_{3}\left(x\right)=\left[1-2r\left(x\right)\right]\left[\frac{1}{1-x}+1-x\right],\cr
 &  & r\left(x\right)=\frac{x}{x_{0}\left(1-x\right)},\cr
 &  & D_{0}\left(x_{0}\right)=\left(1-\frac{4}{x_{0}}-\frac{8}{x_{0}^{2}}\right)\ln\left(1+x_{0}\right)+\frac{1}{2}+\frac{8}{x_{0}}-\frac{1}{2\left(1+x_{0}\right)^{2}},\cr
 &  & D_{1}\left(x_{0}\right)=\left(1+\frac{2}{x_{0}}\right)\ln\left(1+x_{0}\right)-\frac{5}{2}+\frac{1}{1+x_{0}}-\frac{1}{2\left(1+x_{0}\right)^{2}},
\eea
$ d N\left(y\right)/ d y$ and
$\xi_{i}\left(y\right)$ are obtained by replacing $\pcal_{e}$,
$\pcal_{\gamma}$ and $\pcal_{t}$ by $\pcal_{\tilde{e}}$,
$\pcal_{\tilde{\gamma}}$ and $\pcal_{\tilde{t}}$. Here
$\pcal_{e}$ and $\pcal_{\tilde{e}}$ is the longitudinal-polarization
of the incoming electron (positron), respectively, with the definition
as
\beq
\pcal_{e} = \frac{N_{+}-N_{-}}{N_{+}+N_{-}},\qquad
\pcal_{\tilde{e}} = -\frac{N_{+}-N_{-}}{N_{+}+N_{-}},
\eeq
where $N_\pm$ respectively denote the number of electrons or positrons with positive
and negative helicities. $\pcal_{\gamma}$ and $\pcal_{\tilde{\gamma}}$
represent the average helicities of the initial-laser-photons, and
$\pcal_{t}$ and $\pcal_{\tilde{t}}$ represents the maximum
average linear-polarization of the initial-laser-photons. The azimuthal
angle $\varphi$ is defined in the same way as in~\cite{Ginzburg:1981vm}.
We note that $\pcal_{\gamma,t}$ and $\pcal_{\tilde{\gamma},\tilde{t}}$
obey 
\beq
0 \leq \pcal_{\gamma}^{2}+\pcal_{t}^{2}\leq1,\qquad
0 \leq \pcal_{\tilde{\gamma}}^{2}+\pcal_{\tilde{t}}^{2}\leq1.
\eeq
In Fig.~\ref{fig:photonPDF} we show the effective photon spectrum
distributions as a function of $x$ with the polarization parameter
sets listed in Table~\ref{tab:polarizationparameters}. It is clearly
shown that colliding like-handed electrons and photons results in
a flat distribution of backscattered photons while colliding oppositely
handed electrons and photons results in a peaked distribution of backscattered
photons. In both cases the resulting photons are highly polarized~\cite{Ginzburg:1981vm}. 

\section{Helicity Amplitudes~\label{sec:Helicity-notation}}

\subsection{Notation}

In this appendix we briefly summarize our method for calculating the
helicity amplitudes. The method breaks down the algebra of four-dimensional
Dirac spinors and matrices into equivalent two-dimensional ones. In
the Weyl basis, Dirac spinors have the form
\beq
\left(\begin{array}{c}
\psi_{+}\\
\psi_{-}\end{array}\right),
\eeq
where for fermions 
\beq
\psi_{\pm}=\left\{ \begin{array}{cc}
u_{\pm}^{(\lambda=+)}=\omega_{\pm}\chi_{1/2} & ,\\
u_{\pm}^{(\lambda=-)}=\omega_{\mp}\chi_{-1/2} & ,\end{array}\right.\label{eq:fermion}
\eeq
and anti-fermions
\beq
\psi_{\pm}=\left\{ \begin{array}{cc}
v_{\pm}^{(\lambda=1)}=\pm\omega_{\mp}\chi_{-1/2} & ,\\
v_{\pm}^{(\lambda=-)}=\mp\omega_{\pm}\chi_{1/2} & ,\end{array}\right.\label{eq:antifermion}
\eeq
with $\omega_\lambda=\sqrt{E+\lambda\left|\vec{p}\right|};~\lambda=\pm1$, where $E$ and
$\vec{p}$ are the energy and momentum of the fermion, respectively.
The $\chi_{\lambda/2}$'s are eigenvectors of the helicity operator
\beq
h=\hat{p}\cdot\sigma,\,\,{\rm with}\hat{p}=\vec{p}/\left|\vec{p}\right|,
\eeq
where eigenvalue $\lambda=1$ is for ``spin-up'' and $\lambda=-1$
is for ``spin-down'' fermion.
\beq
\chi_{1/2}\equiv\left|\hat{p}+\right\rangle =\left(\begin{array}{c}
\cos\theta/2\\
e^{i\phi}\sin\theta/2\end{array}\right),\,\,\chi_{-1/2}\equiv\left|\hat{p}-\right\rangle =\left(\begin{array}{c}
-e^{i\phi}\sin\theta/2\\
\cos\theta/2\end{array}\right),\label{eq:braket}
\eeq
where we introduce the shorthand notations $\left|\hat{p},\pm\right\rangle $
for $\chi_{\pm1/2}$. We can further simplify (\ref{eq:fermion}, \ref{eq:antifermion})
using the notation
\beq
u_{\pm}(\lambda) = \omega_{\pm\lambda}\left|\hat{p},\lambda\right\rangle ,\qquad
v_{\pm}(\lambda) = \pm\lambda\omega_{\mp\lambda}\left|\hat{p},-\lambda\right\rangle ,
\eeq

Gamma matrices in the Weyl basis have the form
\beq
\gamma^{0}=\left(\begin{array}{cc}
0 & 1\\
1 & 0\end{array}\right),\,\,\,\gamma^{j}=\left(\begin{array}{cc}
0 & -\sigma_{j}\\
\sigma_{j} & 0\end{array}\right),\,\,\,\gamma^{5}=\left(\begin{array}{cc}
1 & 0\\
0 & -1\end{array}\right),
\eeq
where $\sigma_{j}$ are the Pauli $2\times2$ spin matrices. In the
Weyl basis, $\not\! p$ takes the form
\beq
\not\! p\equiv p_{\mu}\gamma^{\mu}=\left(\begin{array}{cc}
0 & p_{0}+\vec{\sigma}\cdot\vec{p}\\
p_{0}-\vec{\sigma}\cdot\vec{p} & 0\end{array}\right)\equiv\left(\begin{array}{cc}
0 & \not\! p_{+}\\
\not\! p_{-} & 0\end{array}\right)\equiv p_{\mu}\left(\begin{array}{cc}
0 & \gamma_{+}^{\mu}\\
\gamma_{-}^{\mu} & 0\end{array}\right)
\eeq
where 
\beq
\gamma_{\pm}^{\mu}=(1,\mp\vec{\sigma}).
\eeq

\subsection{$Wtb$ operators}

\begin{figure}
\includegraphics[scale=0.40]{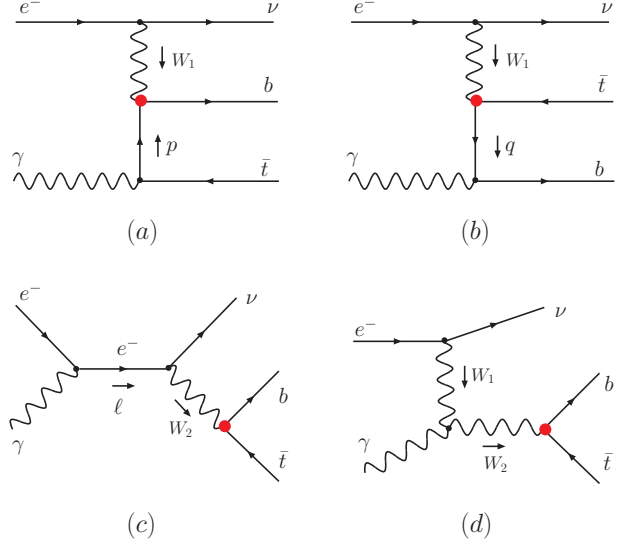}

\caption{Feynman diagrams of process $e^{-}\gamma\to\nu b\bar{t}$ which involve
effective $W-t-b$ couplings ($\fl$ and $\fr$) denoted by the red
circle. The arrow lines beside the propagators indicate the momentum
assignments used in the helicity amplitude calculation.\label{fig:feyndiag_wtb_em}}
\end{figure}

For the process $e^{-}\gamma\to\nu b\bar{t}$, the Feynman diagram
involving the effective $W-t-b$ couplings are shown in Fig.~\ref{fig:feyndiag_wtb_em}.
Unitary gauge is adopted in our calculation. We denote the helicity
amplitude as $\mcal_{Wtb}\left(\lambda_{\gamma},\lambda_{b},\lambda_{\bar{t}}\right)$
where $\lambda_{\gamma}$, $\lambda_{b}$ and $\lambda_{\bar{t}}$
is the helicity of $\gamma$, $b$ and $\bar{t}$, respectively. Below
we give the the helicity amplitudes of process $e^{-}\gamma\to\nu b\bar{t}$
which are induced by the effective $Wtb$ coupling defined in (\ref{eq:wtb}):
\bea
\mcal_{Wtb}\left(\lambda_{\gamma},\lambda_{b},\lambda_{\bar{t}}\right) & = & \left(i\frac{g}{\sqrt{2}}\right)^{2}\left(ie\right)\left(\omega_{+}^{e}\omega_{+}^{\nu}\right)\cr
 & \times & \Biggl\{\frac{Q_{\bar{t}}}{\left(p_{W_{1}}^{2}-m_{W}^{2}+im_{W}\Gamma_{W}\right)\left(p^{2}-m_{t}^{2}\right)}\mcal_{Wtb}^{a}\left(\lambda_{\gamma},\lambda_{b},\lambda_{\bar{t}}\right)\cr
 &  & +\frac{Q_{b}}{\left(p_{W_{1}}^{2}-m_{W}^{2}+im_{W}\Gamma_{W}\right)\left(q^{2}-m_{b}^{2}\right)}\mcal_{Wtb}^{b}\left(\lambda_{\gamma},\lambda_{b},\lambda_{\bar{t}}\right)\cr
 &  & +\frac{1}{\left(p_{W_{2}}^{2}-m_{W}^{2}+im_{W}\Gamma_{W}\right)\ell^{2}}\mcal_{Wtb}^{c}\left(\lambda_{\gamma},\lambda_{b},\lambda_{\bar{t}}\right)\cr
 &  & +\frac{1}{\left(p_{W_{1}}^{2}-m_{W}^{2}+im_{W}\Gamma_{W}\right)\left(p_{W_{2}}^{2}-m_{W}^{2}+im_{W}\Gamma_{W}\right)}\mcal_{Wtb}^{d}\left(\lambda_{\gamma},\lambda_{b},\lambda_{\bar{t}}\right)\Biggr\}
\cr&&
\eea
where 
\bea
\mcal_{Wtb}^{a}\left(\lambda_{\gamma},\lambda_{b},\lambda_{\bar{t}}\right) & = & 2\fl\left[\omega_{-\!\lambda_{b}}\left(-\!\lambda_{\bar{t}}\,\omega_{\lambda_{\bar{t}}}\right)\right]\BSSK{e}{+}{p}{-}{\epsilon}{+}{\bar{t}}{-\!\lambda_{\bar{t}}}\BK{b}{\lambda_{b}}{\nu}{+} \cr
 & + & 2\fr\left[\omega_{\lambda_{b}}\left(\lambda_{\bar{t}}\,\omega_{-\!\lambda_{\bar{t}}}\right)\right]\BSSK{\nu}{-}{p}{+}{\epsilon}{-}{\bar{t}}{-\!\lambda_{\bar{t}}}\BK{b}{\lambda_{b}}{e}{-} \cr
 & + & 2\fl\, m_{t}\left[\omega_{-\!\lambda_{b}}\left(\lambda_{\bar{t}}\,\omega_{-\!\lambda_{\bar{t}}}\right)\right]\BSK{e}{+}{\epsilon}{-}{\bar{t}}{-\!\lambda_{\bar{t}}}\BK{b}{\lambda_{b}}{\nu}{+} \cr
 & + & 2\fr\, m_{t}\left[\omega_{\lambda_{b}}\left(-\!\lambda_{\bar{t}}\,\omega_{\lambda_{\bar{t}}}\right)\right]\BSK{\nu}{-}{\epsilon}{+}{\bar{t}}{-\!\lambda_{\bar{t}}}\BK{b}{\lambda_{b}}{e}{-},\label{eq:ME_a}
\eea

\bea
\mcal_{Wtb}^{b}\left(\lambda_{\gamma},\lambda_{b},\lambda_{\bar{t}}\right) & = & 2\fl\left[\omega_{-\lambda_{b}}\left(-\!\lambda_{\bar{t}}\,\omega_{\lambda_{\bar{t}}}\right)\right]\BSSK{b}{\lambda_{b}}{\epsilon}{+}{q}{-}{\nu}{+}\BK{e}{+}{\bar{t}}{-\!\lambda_{\bar{t}}} \cr
 & + & 2\fr\left[\omega_{\lambda_{b}}\left(\lambda_{\bar{t}}\,\omega_{-\!\lambda_{\bar{t}}}\right)\right]\BSSK{b}{\lambda_{b}}{\epsilon}{-}{q}{+}{e}{-}\BK{\nu}{-}{\bar{t}}{-\!\lambda_{\bar{t}}} \cr
 & + & 2\fl\, m_{b}\left[\omega_{\lambda_{b}}\left(-\!\lambda_{\bar{t}}\,\omega_{\lambda_{\bar{t}}}\right)\right]\BSK{b}{\lambda_{b}}{\epsilon}{-}{\nu}{+}\BK{e}{+}{\bar{t}}{-\!\lambda_{\bar{t}}} \cr
 & + & 2\fr\, m_{b}\left[\omega_{-\lambda_{b}}\left(\lambda_{\bar{t}}\,\omega_{-\!\lambda_{\bar{t}}}\right)\right]\BSK{b}{\lambda}{\epsilon}{+}{e}{-}\BK{\nu}{-}{\bar{t}}{-\!\lambda_{\bar{t}}},\label{eq:ME_b}
\eea

\bea
\mcal_{Wtb}^{c}\left(\lambda_{\gamma},\lambda_{b},\lambda_{\bar{t}}\right) & = & 2\fl\left[\omega_{-\lambda_{b}}\left(-\!\lambda_{\bar{t}}\,\omega_{\lambda_{\bar{t}}}\right)\right]\BK{b}{\lambda_{b}}{\nu}{+}\BSSK{e}{+}{\epsilon}{-}{\ell}{+}{\bar{t}}{-\!\lambda_{\bar{t}}} \cr
 & + & 2\fr\left[\omega_{\lambda_{b}}\left(\lambda_{\bar{t}}\,\omega_{-\!\lambda_{\bar{t}}}\right)\right]\BK{\nu}{-}{\bar{t}}{-\!\lambda_{\bar{t}}}\BSSK{b}{\lambda_{b}}{\ell}{-}{\epsilon}{+}{e}{-}\cr
 & - & \frac{\fl}{m_{W}^{2}}\biggl[m_{b}\,\omega_{\lambda_{b}}\left(-\!\lambda_{\bar{t}}\,\omega_{\lambda_{\bar{t}}}\right)-m_{t}\,\omega_{-\!\lambda_{b}}\left(\lambda_{\bar{t}}\omega_{-\!\lambda_{\bar{t}}}\right)\biggr] \cr
 &  & \,\,\,\,\,\,\,\,\,\,\,\times\BK{b}{\lambda_{b}}{\bar{t}}{-\!\lambda_{\bar{t}}}\left\langle \nu,-|\not\! p_{W_{2}+}\not\!\ell_{-}\not\!\epsilon_{+}|e,-\right\rangle  \cr
 & - & \frac{\fr}{m_{W}^{2}}\biggl[m_{b}\,\omega_{-\!\lambda_{b}}\left(\lambda_{\bar{t}}\,\omega_{-\!\lambda_{\bar{t}}}\right)-m_{t}\,\omega_{\lambda_{b}}\left(-\!\lambda_{\bar{t}}\omega_{\lambda_{\bar{t}}}\right)\biggr] \cr
 &  & \,\,\,\,\,\,\,\,\,\,\,\times\BK{b}{\lambda_{b}}{\bar{t}}{-\!\lambda_{\bar{t}}}\left\langle \nu,-|\not\! p_{W_{2}+}\not\!\ell_{-}\not\!\epsilon_{+}|e,-\right\rangle ,\label{eq:ME_c}
\eea

\bea
\mcal_{Wtb}^{d}\left(\lambda_{\gamma},\lambda_{b},\lambda_{\bar{t}}\right) & = & -\fl\left[\omega_{-\lambda_{b}}\left(-\!\lambda_{\bar{t}}\,\omega_{\lambda_{\bar{t}}}\right)\right] \cr
 &  & \,\,\,\,\,\,\,\,\times\biggr[2\BK{b}{\lambda_{b}}{\nu}{+}\BK{e}{+}{\bar{t}}{-\!\lambda_{\bar{t}}}\left(p_{W_{1}}+p_{W_{2}}\right)\cdot\epsilon  \cr
 &  & \,\,\,\,\,\,\,\,\,\,\,\,\,\,-\BSK{\nu}{-}{\left(\not\! p_{W_{2}}+\not\! p_{\gamma}\right)}{+}{e}{-}\BSK{b}{\lambda_{b}}{\epsilon}{+}{\bar{t}}{-\!\lambda_{\bar{t}}} \cr
 &  & \,\,\,\,\,\,\,\,\,\,\,\,\,\,+\BSK{\nu}{-}{\epsilon}{+}{e}{-}\BSK{b}{\lambda_{b}}{\left(\not\! p_{\gamma}-\not\! p_{W_{1}}\right)}{+}{\bar{t}}{-\!\lambda_{\bar{t}}}\biggr] \cr
 & - & \fr\left[\omega_{\lambda_{b}}\left(\lambda_{\bar{t}}\,\omega_{-\!\lambda_{\bar{t}}}\right)\right] \cr
 &  & \,\,\,\,\,\,\,\,\times\biggr[2\BK{\nu}{-}{\bar{t}}{-\!\lambda_{\bar{t}}}\BK{b}{\lambda_{b}}{e}{-}\left(p_{W_{1}}+p_{W_{2}}\right)\cdot\epsilon \cr
 &  & \,\,\,\,\,\,\,\,\,\,\,\,\,\,-\BSK{\nu}{-}{\left(\not\! p_{W_{2}}+\not\! p_{\gamma}\right)}{+}{e}{-}\BSK{b}{\lambda_{b}}{\epsilon}{-}{\bar{t}}{-\!\lambda_{\bar{t}}} \cr
 &  & \,\,\,\,\,\,\,\,\,\,\,\,\,\,+\BSK{\nu}{-}{\epsilon}{+}{e}{-}\BSK{b}{\lambda_{b}}{\left(\not\! p_{\gamma}-\not\! p_{W_{1}}\right)}{-}{\bar{t}}{-\!\lambda_{\bar{t}}}\biggr] \cr
 & + & \fl\,\frac{2p_{e}\cdot p_{\gamma}}{m_{W}^{2}}\biggl[m_{b}\,\omega_{\lambda_{b}}\left(-\!\lambda_{\bar{t}}\,\omega_{\lambda_{\bar{t}}}\right)-m_{t}\,\omega_{-\!\lambda_{b}}\left(\lambda_{\bar{t}}\omega_{-\!\lambda_{\bar{t}}}\right)\biggr] \cr
 &  & \,\,\,\,\,\,\,\,\,\,\,\,\,\,\,\,\,\,\,\,\,\,\,\,\,\,\,\times\BK{b}{\lambda_{b}}{\bar{t}}{-\!\lambda_{\bar{t}}}\BSK{\nu}{-}{\epsilon}{+}{e}{-} \cr
 & + & \fr\,\frac{2p_{e}\cdot p_{\gamma}}{m_{W}^{2}}\biggl[m_{b}\,\omega_{-\!\lambda_{b}}\left(\lambda_{\bar{t}}\,\omega_{-\!\lambda_{\bar{t}}}\right)-m_{t}\,\omega_{\lambda_{b}}\left(-\!\lambda_{\bar{t}}\omega_{\lambda_{\bar{t}}}\right)\biggr] \cr
 &  & \,\,\,\,\,\,\,\,\,\,\,\,\,\,\,\,\,\,\,\,\,\,\,\,\,\,\,\times\BK{b}{\lambda_{b}}{\bar{t}}{-\!\lambda_{\bar{t}}}\BSK{\nu}{-}{\epsilon}{+}{e}{-},\label{eq:ME_d}
\eea
where $\lambda_{\gamma},\lambda_{b},\lambda_{\bar{t}}=\mp1$ corresponds
to left-handed and right-handed helicity respectively. The momentum
of the propagators are defined as follows: $p_{W_{1}}=p_{e}-p_{\nu}$,
$p_{W_{2}}=p_{b}+p_{\bar{t}}$, $p=p_{\gamma}-p_{b}$, $q=p_{\gamma}-p_{\bar{t}}$, 
$\ell=p_{\gamma}+p_{e^{-}}$, cf. Fig.~\ref{fig:feyndiag_wtb_em},
and $\epsilon\left(\lambda_{\gamma}\right)$ is the polarization vector
of the incoming photon. The weight factor $\omega_{\lambda_{i}}$
depends on the helicity ($\lambda_{i}$) of the fermion $i$:
\beq
\omega_{\lambda_{i}}\equiv\sqrt{E_{i}+\lambda_{i}\left|\vec{p_{i}}\right|},
\eeq
where $E_{i}$ and $\vec{p}_{i}$ is the energy and momentum of the
fermion $i$, respectively.

It is very straightforward to get the SM helicity amplitudes from by setting $\fl=1$, and $\fr=0$.

\subsection{Four-fermion operator}

\begin{figure}
\includegraphics[scale=0.5]{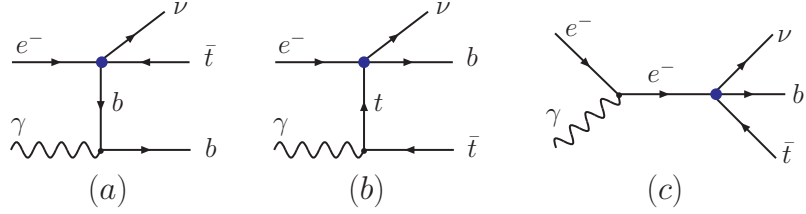}

\caption{Feynman diagrams of process $e^{-}\gamma\to\nu b\bar{t}$ which involve
effective $4$-fermion couplings $\g$ denoted by the blue bulb. \label{fig:feyndiag_wtb_4f}}
\end{figure}

The Feynman diagrams involving effective $4$-fermion couplings are
shown in Fig.~\ref{fig:feyndiag_wtb_4f}. Denoting the helicity amplitude
as $\mcal_{4f}\left(\lambda_{\gamma},\lambda_{b},\lambda_{\bar{t}}\right)$
where $\lambda_{\gamma}$, $\lambda_{b}$ and $\lambda_{\bar{t}}$
is the helicity of $\gamma$, $b$ and $\bar{t}$, respectively, we
obtain the matrix element as follows:
\bea
\mcal_{4f}\left(\lambda_{\gamma},\lambda_{b},\lambda_{\bar{t}}\right) & = & \left(i\g\right)\left(ie\right)\left(\omega_{+}^{e}\omega_{+}^{\nu}\right) \cr
 & \times & \Biggl\{\frac{Q_{\bar{t}}}{p^{2}-m_{t}^{2}}\mcal_{4f}^{a}\left(\lambda_{\gamma},\lambda_{b},\lambda_{\bar{t}}\right)+\frac{Q_{b}}{q^{2}-m_{b}^{2}}\mcal_{4f}^{b}\left(\lambda_{\gamma},\lambda_{b},\lambda_{\bar{t}}\right)\cr
 &  & +\frac{1}{\ell^{2}}\mcal_{4f}^{c}\left(\lambda_{\gamma},\lambda_{b},\lambda_{\bar{t}}\right)\Biggr\},\label{eq:Hel_4f}
\eea
where
\bea
\mcal_{4f}^{a}\left(\lambda_{\gamma},\lambda_{b},\lambda_{\bar{t}}\right) & = & 2\left[\omega_{-\!\lambda_{b}}\left(-\!\lambda_{\bar{t}}\,\omega_{\lambda_{\bar{t}}}\right)\right]\BSSK{e}{+}{p}{-}{\epsilon}{+}{\bar{t}}{-\!\lambda_{\bar{t}}}\BK{b}{\lambda_{b}}{\nu}{+} \cr
 & + & 2\, m_{t}\left[\omega_{-\!\lambda_{b}}\left(\lambda_{\bar{t}}\,\omega_{-\!\lambda_{\bar{t}}}\right)\right]\BSK{e}{+}{\epsilon}{-}{\bar{t}}{-\!\lambda_{\bar{t}}}\BK{b}{\lambda_{b}}{\nu}{+},\label{eq:ME_a_4f}
\eea

\bea
\mcal_{4f}^{b}\left(\lambda_{\gamma},\lambda_{b},\lambda_{\bar{t}}\right) & = & 2\left[\omega_{-\lambda_{b}}\left(-\!\lambda_{\bar{t}}\,\omega_{\lambda_{\bar{t}}}\right)\right]\BSSK{b}{\lambda_{b}}{\epsilon}{+}{q}{-}{\nu}{+}\BK{e}{+}{\bar{t}}{-\!\lambda_{\bar{t}}} \cr
 & + & 2\, m_{b}\left[\omega_{\lambda_{b}}\left(-\!\lambda_{\bar{t}}\,\omega_{\lambda_{\bar{t}}}\right)\right]\BSK{b}{\lambda_{b}}{\epsilon}{-}{\nu}{+}\BK{e}{+}{\bar{t}}{-\!\lambda_{\bar{t}}},\label{eq:ME_b_4f}
\eea

\bea
\mcal_{4f}^{c}\left(\lambda_{\gamma},\lambda_{b},\lambda_{\bar{t}}\right) & = & 2\left[\omega_{-\lambda_{b}}\left(-\!\lambda_{\bar{t}}\,\omega_{\lambda_{\bar{t}}}\right)\right]\BK{b}{\lambda_{b}}{\nu}{+}\BSSK{e}{+}{\epsilon}{-}{\ell}{+}{\bar{t}}{-\!\lambda_{\bar{t}}},\label{eq:ME_c_4f}
\eea
where the momentum of propagators is the same as the ones of effective
$Wtb$ couplings. 

\newpage
\bibliography{reference}

\end{document}

%% file: mylatex.tex
\usepackage{bbm}                                                  

\newcommand{\nc}{\newcommand}
\nc{\beq}{\begin{equation}}  \nc{\eeq}{\end{equation}}
\nc{\bea}{\begin{eqnarray}}  \nc{\eea}{\end{eqnarray}}
\nc{\baa}{\begin{array}}     \nc{\eaa}{\end{array}}
\nc{\bit}{\begin{itemize}}   \nc{\eit}{\end{itemize}}
\nc{\ben}{\begin{enumerate}} \nc{\een}{\end{enumerate}}
\nc{\bce}{\begin{center}}    \nc{\ece}{\end{center}}
\nc{\bpm}{\begin{pmatrix}}   \nc{\epm}{\end{pmatrix}}
\nc{\bvt}{\begin{verbatim}}  \nc{\evt}{\end{verbatim}}
\def\half{\frac12}	

\def\to{\rightarrow}

\def\gesim{\,{\raise-3pt\hbox{$\sim$}}\!\!\!\!\!{\raise2pt\hbox{$>$}}\,}
\def\lesim{\,{\raise-3pt\hbox{$\sim$}}\!\!\!\!\!{\raise2pt\hbox{$<$}}\,}
\def\boldoverdot{\,{\raise6pt\hbox{\bf.}\!\!\!\!\>}}

\def\bcal{{\cal B}}

\def\fcal{{\cal F}}
\def\gcal{{\cal G}}

\def\lcal{{\cal L}}
\def\mcal{{\cal M}}

\def\ocal{{\cal O}}
\def\pcal{{\cal P}}

\def\scal{{\cal S}}

\usepackage{bm}
\def\diag{\hbox{\diag}}
\def\sm{Standard Model}

\def\gev{\hbox{GeV}}
\def\tev{\hbox{TeV}}

\def\vevof#1{\left\langle#1\right\rangle}
\def\doubleundertext#1{
{\undertext{\vphantom{y}#1}}\par\nobreak\vskip-\the\baselineskip\vskip4pt%
\undertext{\hbox to 2in{}}}
\def\inbox#1{\vbox{\hrule\hbox{\vrule\kern5pt
     \vbox{\kern5pt#1\kern5pt}\kern5pt\vrule}\hrule}}
\def\sqr#1#2{{\vcenter{\hrule height.#2pt
      \hbox{\vrule width.#2pt height#1pt \kern#1pt
         \vrule width.#2pt}
      \hrule height.#2pt}}}

\def\today{\ifcase\month\or
  January\or February\or March\or April\or May\or June\or
  July\or August\or September\or October\or November\or December\fi
  \space\number\day, \number\year}
\def\pmb#1{\setbox0=\hbox{#1}%
  \kern-.025em\copy0\kern-\wd0
  \kern.05em\copy0\kern-\wd0
  \kern-.025em\raise.0433em\box0 }
\def\up#1{^{\left( #1 \right) }}
\def\inv#1{\frac1{#1}}
\def\su#1{{SU(#1)}}

\def\sumprime_#1{\setbox0=\hbox{$\scriptstyle{#1}$}
  \setbox2=\hbox{$\displaystyle{\sum}$}
  \setbox4=\hbox{${}'\mathsurround=0pt$}
  \dimen0=.5\wd0 \advance\dimen0 by-.5\wd2
  \ifdim\dimen0>0pt
  \ifdim\dimen0>\wd4 \kern\wd4 \else\kern\dimen0\fi\fi
\mathop{{\sum}'}_{\kern-\wd4 #1}}
\font\eightrm=cmr8